\newcounter{myctr}
\def\myitem{\refstepcounter{myctr}\bibfont\noindent\ifnum\themyctr>9\else\phantom{0}\fi\hangindent17pt\themyctr.\enskip}

\documentclass{ws-ijqi}

\begin{document}

\newcommand{\REV}[1]{\textbf{\color{blue}[[#1]]}}
\newcommand{\GREEN}[1]{\textbf{\color{green}#1}}
\newcommand{\RED}[1]{\textrm{\color{red}#1}}
\newcommand{\rev}[1]{{\color{blue}#1}}

\newcommand{\andy}[1]{ }
\newcommand{\bmsub}[1]{\mbox{\boldmath\scriptsize $#1$}}

\def\R{\mathbb{R}}

\def\bra#1{\langle #1 |}
\def\ket#1{| #1 \rangle}
\def\sinc{\mathop{\text{sinc}}\nolimits}
\def\cV{\mathcal{V}}
\def\cH{\mathcal{H}}
\def\cT{\mathcal{T}}
\def\cM{\mathcal{M}}
\def\cN{\mathcal{N}}
\def\CW{\mathcal{W}}
\def\e{\mathrm{e}}
\def\ii{\mathrm{i}}
\def\d{\mathrm{d}}
\renewcommand{\Re}{\mathop{\text{Re}}\nolimits}
\newcommand{\tr}{\mathop{\text{Tr}}\nolimits}

\markboth{Francesco V. Pepe} {Multipartite Entanglement and
Few-Body Hamiltonians}

%%%%%%%%%%%%%%%%%%%%% Publisher's Area please ignore %%%%%%%%%%%%%%
\catchline{}{}{}{}{}
%%%%%%%%%%%%%%%%%%%%%%%%%%%%%%%%%%%%%%%%%%%%%%%%%%%%%%%%%%%%%%%%%%%

\title{MULTIPARTITE ENTANGLEMENT AND FEW-BODY HAMILTONIANS}

\author{FRANCESCO V. PEPE}

\address{Dipartimento di Fisica, Universit\`{a} di Bari and INFN - Sezione di Bari, Via Amendola 173\\
Bari, 70126, Italy\\
francesco.pepe@ba.infn.it}

\maketitle

\begin{history}
\received{Day Month Year}
\revised{Day Month Year}
%\accepted{Day Month Year}
%\comby{(xxxxxxxxxx)}
\end{history}

\begin{abstract}
We investigate the possibility to obtain higly
multipartite-entangled states as {\it nondegenerate} eigenstates
of Hamiltonians that involve only short-range and few-body
interactions. We study small-size systems (with a number of qubits
ranging from three to five) and search for Hamiltonians with a
Maximally Multipartite Entangled State (MMES) as a nondegenerate
eigenstate. We then find conditions, including bounds on the
number of coupled qubits, to build a Hamiltonian with a
Greenberger-Horne-Zeilinger (GHZ) state as a nondegenerate
eigenstate. We finally comment on possible applications.
\end{abstract}


\section{Introduction}

Multipartite entanglement is an inherently quantum phenomenon
whose features are attracting increasing attention lately. While
in the bipartite case different mathematical definitions are
physically equivalent,\cite{entanglement,entanglementrev,h4} a
unique characterization of multipartite quantum correlations does
not exist. Alternative proposals, highlighting different aspects
of this phenomenon are
possible.\cite{multipart1,multipart2,multipart3,multipart4,Bergou,scott}
Moreover, multipartite entanglement is characterized by features
appearing also in other areas of physics, such as complexity and
frustration.\cite{nielsen1,Mejìa-Monasterio,NJP,Marzolino} The
interest in multipartite entanglement has been motivated by
possible applications in quantum enhanced tasks,\cite{Adesso} but
also by genuine foundational aspects.

It was proposed by P.~Facchi {\it et al.}~in Ref.~\refcite{MMES}
that the multipartite entanglement of a system of qubits can be
characterized in terms of the distribution of bipartite
entanglement, quantified by a proper measure (e.g.~purity), over
all possible bipartitions of the system. This leads to the notion
of ``maximally multipartite entangled states" (MMES), as those
states for which average purity over all {\it balanced}
bipartitions is minimal. Another paradigmatic example of
multipartite-entangled states are the $n$-qubit
Greenberger-Horne-Zeilinger states (GHZ),\cite{greenberger} whose
quantum correlations are purely
$n$-partite,\cite{cirac,multipart1} in the sense that they do not
retain any entanglement as one traces out one of the qubits.

By their very definition, highly multipartite-entangled states
exhibit nonlocal correlations which are both strong and
\emph{distributed} among different bipartitions. This naturally
leads to the following question: can MMES or GHZ be obtained as
nondegenerate ground states of Hamiltonians that only involve
\emph{local} interactions?\cite{buzek,loss,loss1,verrucchi} In the
context of spin systems, ``local" means both few-body and
nearest-neighbor. If this were impossible, one could soften the
requirement and ask whether one can find Hamiltonians containing
up to two-body interaction terms, whose (nondegenerate) eigenstate
is a MMES or GHZ state. These problems were tackled in
Refs.~\refcite{eigenmmes} and \refcite{ghz}. While the answer to
the ground-state question is negative, it is instead possible to
find local Hamiltonians with multipartite-entangled nondegenerate
eigenstates. In this paper, we will analyze the features of
Hamiltonians with MMES eigenstates for the cases $n=3,4\text{ and
}5$ qubits and of a family of Hamiltonians characterized by a GHZ
eigenstate for any number of qubits.

This paper is organized as follows. In Section 2 we briefly
outline the entanglement features of MMES and GHZ states,
highlighting analogies and differences. In Section 3 we outline
the problems which arise in finding a multipartite-entangled
eigenstate of a local Hamiltonian, verify the impossibility to
have a nondegenerate MMES or GHZ ground state and study a simple
and relevant counterexample, involving W states. In Section 4,
two-body and nearest-neighbor Hamiltonians with nondegenerate MMES
eigenstate in the cases of $n=3,4,5$ qubits are analyzed, while in
Section 5 a set of Hamiltonians which are ``as local as possible''
and have a GHZ nondegenerate eigenstate is studied. We conclude by
discussing possible applications and uses of this kind of
operators to generate entanglement.

\section{Multipartite entanglement: MMES and GHZ states}

Maximally Multipartite Entangled States have been defined in
Ref.~\refcite{MMES}, following the idea that multipartite
entanglement is strong if {\it bipartite} correlations are large
and very well-distributed among all different bipartitions. Let us
consider a system of $n$ qubits in a pure state $\ket{\psi}$ and
an arbitrary bipartition in two subsystems $(A,\bar{A})$,
including respectively $n_A$ and $n_{\bar{A}}$ qubits, with
$n_A+n_{\bar{A}}=n$ and $n_A\leq n_{\bar{A}}$. Bipartite
entanglement between $A$ and $\bar{A}$ can be quantified by purity
\begin{equation}
\pi_A (\ket{\psi}) = \mathrm{Tr}_A \left( \rho_A^2 \right), \quad
\text{with} \quad \rho_A = \mathrm{Tr}_{\bar{A}} \left( \ket{\psi}
\bra{\psi} \right).
\end{equation}
Here, $\rho_A$ is the reduced density matrix of subsystem $A$ and
$\mathrm{Tr}_A$ ($\mathrm{Tr}_{\bar{A}}$) denotes the trace over
the Hilbert space of subsystem $A$ ($\bar{A}$). Purity is bound in
the interval
\begin{equation}\label{eq:boundpur}
\frac{1}{2^{n_A}} \leq \pi_{A} (\ket{\psi}) \leq 1.
\end{equation}
The maximum is attained by states which are factorized with
respect to the given bipartition. On the other side, maximally
entangled states, whose reduced density matrix $\rho_A$ is
maximally mixed with respect to the bipartition, saturate the
minimum of purity.

Multipartite entanglement can be described in terms of the {\it
distribution of purity over all possible bipartitions} of the
system. Clearly, the number of functions one needs for a complete
characterization of multipartite quantum correlations scales
exponentially with the size of the system. The simplest quantity
which can be used to characterize multipartite entanglement is an
average value of purity. In Ref. \refcite{MMES} the {\it potential
of multipartite entanglement}, defined as the mean purity over all
{\it balanced} bipartitions with $n_{A}=[n/2]$, was introduced. It
can be expressed as\cite{MMES,multent}
\begin{equation}\label{eq:pme}
\pi_{\mathrm{ME}} (\ket{\psi}) = \frac{1}{C_{n_A}^n}
\sum_{n_A=[n/2]} \pi_A (\ket{\psi}),
\end{equation}
where $C_{n_A}^n$ is the binomial coefficient and the sum runs
over all balanced bipartitions. The potential of multipartite
entanglement inherits the bound \eqref{eq:boundpur}:
\begin{equation}\label{eq:boundpme}
\frac{1}{2^{[n/2]}} \leq \pi_{\mathrm{ME}} (\ket{\psi}) \leq 1.
\end{equation}
The maximum is attained by states which are {\it completely
factorized}, with every qubit in a pure state, while the lower
bound can be saturated only if a state $\ket{\psi}$ is maximally
entangled with respect to {\it all} possible bipartitions of the
system. MMES are generally defined as the minimizers of the
function $\pi_{\mathrm{ME}}$ for a system of $n$ qubits. Actually,
the lower bound $2^{-[n/2]}$ is attained only for a very small set
of qubit numbers, namely $n=3,5,6$ (the case $n=2$ is trivial.) In
these cases, minimizers are called {\it perfect} MMES. In
particular, in the case $n=4$, there exists numerical and
analytical evidence\cite{higuchi1,higuchi2,MMES,wallach} that the
minimal potential of multipartite entanglement is
$\pi_{\mathrm{ME}} = 1/3
> 1/4$.

It is clear by their definition that no unique expression for MMES
states exists, and their form should be obtained by analytical
and/or numerical minimization of $\pi_{\mathrm{ME}}$. This aspects
will be reflected in a lack of homogeneity in the form of the
Hamiltonians with MMES eigenstates, which have few common features
as the number of qubit varies and can also include unusual
interaction terms.

MMES for $n=3$ are equivalent by single-qubit unitary
transformations to the Greenberger-Horne-Zeilinger (GHZ) state
\begin{equation}\label{eq:g3}
\ket{G_+^3} = \frac{1}{\sqrt{2}} \left( \ket{000} + \ket{111}
\right).
\end{equation}
This definition can be generalized in a straightforward way to an
arbitrary number of qubits:
\begin{equation}\label{eq:ghz}
|G_{\pm}^{n}\rangle=\frac{1}{\sqrt{2}}
\left(|0\rangle^{\otimes n}\pm |1\rangle^{\otimes n}\right)
\end{equation}
where $\sigma^z|i\rangle=(-1)^i | i \rangle$ defines the
computational basis, with $i=0,1$ and $\sigma^z$ the third Pauli
matrix. GHZ states represent a generalization of Bell
states.\cite{greenberger} Their entanglement features, as well as
their definition, are very homogeneous as $n$ varies. Since the
purity of a maximally unbalanced bipartition ($n_A=1$) is always
one half, each qubit is maximally entangled to the rest of the
chain. However, $\pi_A=1/2$ for all bipartitions of the system,
including the balanced ones. This means that for $n>3$, GHZ states
are far from being minimizers of the potential of multipartite
entanglement. Another general feature of GHZ states is that if one
of the qubits (say qubit 1 for definiteness) is traced out, the
reduced density matrix is unentangled:
\begin{equation}\label{eq:trace}
\rho_{\{2,3,\dots,n\}} = \mathrm{Tr}_{\{1\}} \left( \ket{
G_{\pm}^{n} } \bra{ G_{\pm}^{n} } \right) \nonumber = \frac{1}{2}
\left( \ket{0_2\dots 0_n} \bra{0_2\dots 0_n} + \ket{1_2\dots 1_n}
\bra{1_2\dots 1_n} \right).
\end{equation}
As we shall see in the following discussion, this aspect is
closely related to the difficulty of obtaining nondegenerate GHZ
eigenstates.

\section{The ground state problem}

In order to understand where the difficulty of finding a
Hamiltonian with an entangled nondegenerate ground state lies, let
us consider an $n$-qubit state $\ket{\psi_n}$ and let us denote
with $H_2^{(n)}$ a local Hamiltonian acting on the $n$-qubit
Hilbert space.

It is not difficult to build a generic Hamiltonian characterized
by $\ket{\psi_n}$ as the nondegenerate ground state, the simplest
example being
\begin{equation}\label{eq:proj}
H = - \ket{\psi_n} \bra{\psi_n}.
\end{equation}
However, the projector appearing in \eqref{eq:proj} generally
contains up to $n$-body operators, which make the resulting
Hamiltonian nonlocal. In order to obtain a local Hamiltonian
$H_2^{(n)}$, one should for example add to \eqref{eq:proj}
suitable Hermitian terms which cancel the nonlocal parts in the
projector and preserve at the same time the eigenstate property of
$\ket{\psi_n}$. This task is easy for states in which no quantum
correlation between qubits is present: the factorized state
$\ket{1}^{\otimes n}$ is the ground state of the one-body
Hamiltonian $\sum_{i=1}^n \sigma_i^z$. When entangled states are
taken into account, problems arise because the two-body matrix
elements of $H_2^{(n)}$ do not capture all the correlations of the
state $\ket{\psi_n}$, which can be complex and encoded in
higher-order correlation functions.

We can analyze how incomplete information on a state included in
the reduced density matrices affects the ground-state problem. For
general theorems on the topic, see
Refs.~\refcite{nielsen1,nielsen2,vandennest}. Let us consider a
local Hamiltonian $H_2^{(n)}$ and its ground state energy
\begin{equation}
E_0 = \min \mathrm{spec} \left( H_2^{(n)} \right),
\end{equation}
with $\mathrm{spec}(O)$ denoting the spectrum of operator $O$. Let
us assume that the state $\ket{\psi_n}$ is a ground state, and
that there exists another state $\ket{\phi_n}$ which is linearly
independent of $\ket{\psi_n}$ and is characterized by the same
two-body reduced density matrices. Since the expectation values of
$H_2^{(n)}$ on a state are completely determined by two-body
reduced density matrices, it follows that
\begin{equation}
\bra{\phi_n} H_2^{(n)} \ket{\phi_n} = \bra{\psi_n} H_2^{(n)}
\ket{\psi_n} = E_0.
\end{equation}
Being $E_0$ the lowest energy level, the above result implies that
$\ket{\psi_n}$ is also an eigenstate, and thus the ground state is
(at least) two-fold degenerate.

In a system of $n=3$ qubits, for example, the MMES state
\eqref{eq:g3} has the same reduced density matrices, as
\begin{equation}
\ket{ G_-^3} = \frac{1}{\sqrt{2}} \left( \ket{000} - \ket{111}
\right).
\end{equation}
Thus, if one of the states $\ket{G_{\pm}^3}$ is a ground state of
a local Hamiltonian, it must be at least twofold degenerate. In
Section 3 we shall find that the degeneracy of a MMES ground state
increases with $n$, as well as the number of states which are
orthogonal to the MMES we are taking into account, but share the
same two-body reduced density matrices. The GHZ case is even
worse, since all the $k$-body reduced density matrices (with
$k<n$) of the states $\ket{G_{\pm}^n}$ in Eq.~\eqref{eq:ghz}
coincide. A Hamiltonian can have a nondegenerate GHZ ground state
{\it only if} it involves $n$-body couplings. As we shall see in
the following section, the increasing number of orthogonal states
sharing the same expectation values accounts also for a greater
difficulty for a MMES to be one of the lowest excited states.

\subsection{A counterexample: W states}

In this subsection we shall show how the problems in finding a
MMES or GHZ ground state of a local Hamiltonian do not arise in
the case of W states. These are defined\cite{cirac} as
superpositions of computational basis states with all but one spin
aligned to the $z$ axis:
\begin{equation}\label{eq:Wn}
\ket{W_n} = \frac{1}{\sqrt{n}} \left( \ket{0_1 0_2 \dots 0_{n-1}
1_n} + \ket{0_1 0_2 \dots 1_{n-1} 0_n} + \dots \ket{1_1 0_2 \dots
0_{n-1} 0_n} \right).
\end{equation}
Though W states are entangled with respect to all bipartitions
like MMES and GHZ states, information on their quantum
correlations is fully encoded in their two-body reduced density
matrices\cite{multipart1,ckwineq}. This is reflected in the
possibility of finding $n$-qubit Hamiltonians with $\ket{W_n}$
ground states in some range of parameters.

An example is given by the class of local XX Hamiltonians with an
external field
\begin{equation}\label{eq:Hb}
H(b) = - \sum_{i=1}^n \left( \sigma_i^x \sigma_{i+1}^x +
\sigma_i^y \sigma_{i+1}^y \right) + b \sum_{i=1}^n \sigma_i^z,
\end{equation}
with $\bm{\sigma}_{n+1}\equiv \bm{\sigma}_1$, made up of a
ferromagnetic Heisenberg coupling in the $xy$ plane and an
interaction with an external field $b$ along the orthogonal $z$
axis. It is possible to check that $\ket{W_n}$ is an eigenstate of
\eqref{eq:Hb} for all $n$ and all values of $b$, with eigenvalue
\begin{equation}
\bra{W_n} H(b) \ket{W_n} = -4 + b (n-2),
\end{equation}
since
\begin{equation}
\sum_{i=1}^n \left( \sigma_i^x \sigma_{i+1}^x + \sigma_i^y
\sigma_{i+1}^y \right) \ket{W_n} = 2 \sum_{i=1}^n \left(
\sigma_i^+ \sigma_{i+1}^- + \sigma_i^- \sigma_{i+1}^+ \right)
\ket{W_n} = 4 \ket{W_n}
\end{equation}
and
\begin{equation}
\sum_{i=1}^n \sigma_i^z \ket{W_n} = \left[ (n-1)-1 \right]
\ket{W_n} = (n-2) \ket{W_n}.
\end{equation}
The Hamiltonian can be analytically diagonalized for $n$ ranging
from $3$ to $6$. Is is also possible in these cases to find a
range of parameters $b$ in which $\ket{W_n}$ is the {\it
nondegenerate ground state}. These ranges are reported in Table
\ref{tab:W}.
\begin{table}\label{tab:W}
\tbl{Relevant properties of $\ket{W_n}$ \eqref{eq:Wn} as an
eigenstate of $H(b)$, defined in \eqref{eq:Hb}. In the third
column, the range in which it is the ground state is indicated.
$\ket{W_n}$ is degenerate at the extrema and nondengenerate inside
the intervals.} {
\begin{tabular}{@{}ccc@{}} \toprule $n$ &
Eigenvalue of $\ket{W_n}$ & Range for $\ket{W_n}$ ground state  \\
\colrule
3 & $-4+b$ & $-2 \leq b \leq 0$ \\
4 & $-4+2b$ & $-2 \leq b \leq - 2 (\sqrt{2} - 1)$ \\
5 & $-4+3b$ & $-2 \leq b \leq - (\sqrt{5} - 1)$ \\
6 & $-4+4b$ & $-2 \leq b \leq - 2 (\sqrt{3} - 1)$ \\ \botrule
\end{tabular}}
\end{table}
Numerical evidence shows that $\ket{W_n}$ ceases to be a
nondegenerate ground state for $n\geq 7$. This, of course, does
not mean that it would not be possible to find Hamiltonians which
are more suitably tailored (and perhaps more complicated) than
\eqref{eq:Hb}, and satisfy this property even for $n>6$.

\section{Local Hamiltonians with MMES eigenstates}

In this section, we will study how is it possible to obtain MMES
eigenstates of two-body Hamiltonians in the cases of $n=3,4,5$
qubits. The focus will be mainly on the features of special
Hamiltonians, since a discussion on the most general operators
with MMES eigenstates is given in Ref.~\refcite{eigenmmes}.

\subsection{$n=3$}

In the case of three qubits, MMES are equivalent by single-qubit
unitary operations to the GHZ state. Since the latter has a simple
and symmetric form, let us choose to study its properties, with no
significant loss of generality:
\begin{equation}
\ket{M_3} \equiv \ket{G_+^3} = \frac{1}{\sqrt{2}} \left( \ket{000}
+ \ket{111} \right).
\end{equation}
Since $\pi_{A}=1/2$ for each bipartition, $\ket{M_3}$ saturates
the lower bound in \eqref{eq:boundpme} and is therefore a perfect
MMES. Moreover, the GHZ form of the state implies that it shares
two-body reduced density matrices with its antisymmetric
counterpart $\ket{G_-^3}$, meaning that it can never be the
nondegenerate ground state of a two-body Hamiltonian.

\begin{figure}
\centering
\includegraphics[width=0.8\textwidth]{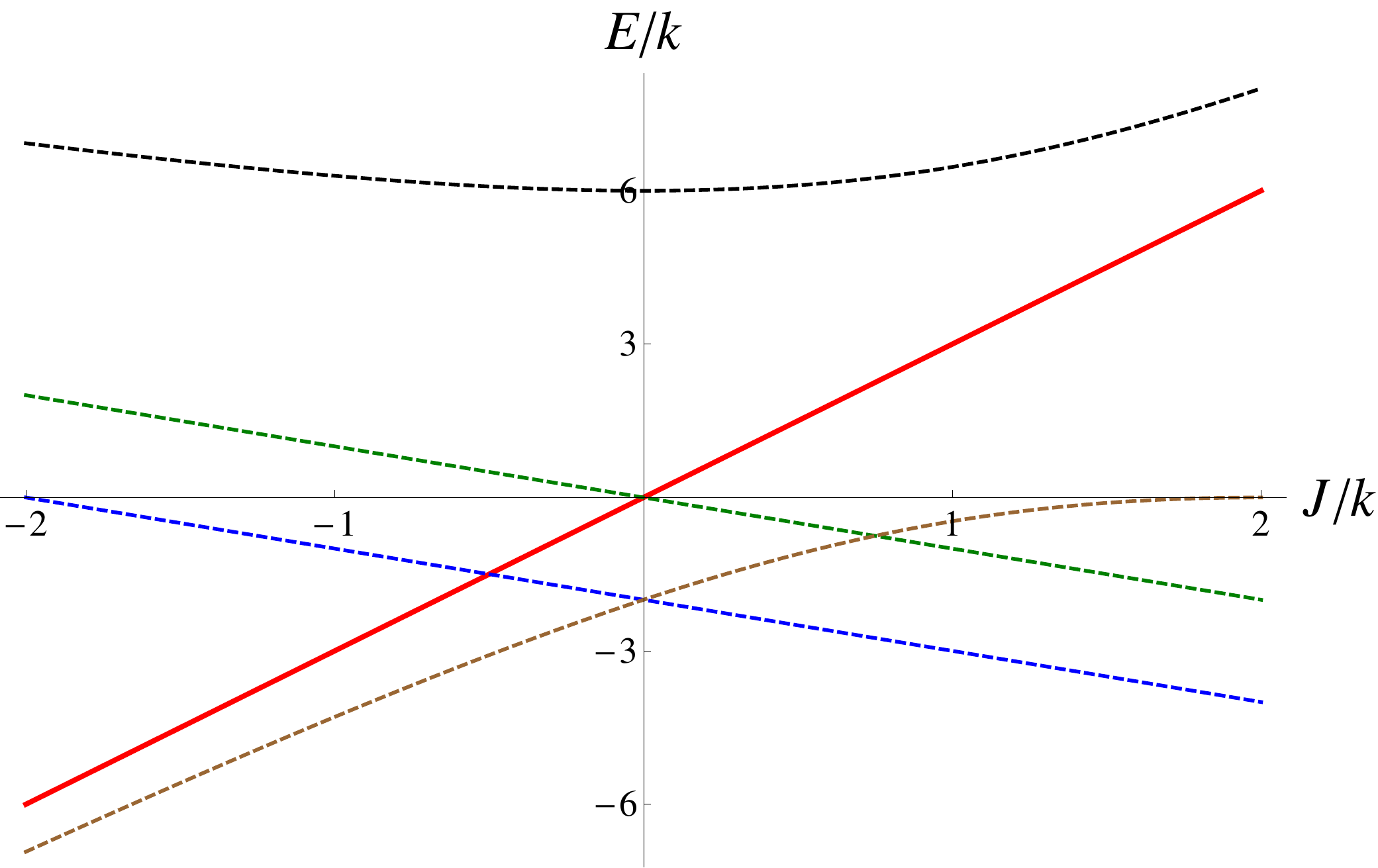}
\caption{Spectrum of the Hamiltonian $H_3(J,k)$ in Eq.
\eqref{eq:H3Jk} as a function of the reduced parameter $J/k$. The
solid (red) line represents the eigenvalue $3J$ associated to the
eigenstate $\ket{M_3}$, while dashed lines are the other energy
levels. Notice that the eigenvalue $-(J+2k)$ is twofold degenerate
and the eigenvalue $-J$ is threefold
degenerate.}\label{fig:spectrum3}
\end{figure}

The most general two-body Hamiltonian in the $n=3$ case, which is
{\it a fortiori} also nearest-neighbor, reads, apart from an
irrelevant constant
\begin{equation}\label{eq:H3}
H = \sum_{i=1}^3 \left( \sum_{\alpha,\beta=x,y,z}
\gamma_{i}^{\alpha\beta} \sigma_i^{\alpha} \sigma_{i+1}^{\beta} +
\sum_{\alpha=x,y,z} h_i^{\alpha} \sigma_{i}^{\alpha} \right),
\end{equation}
with the convention $\bm{\sigma}_4\equiv \bm{\sigma}_1$. A first
step consists in determining which conditions must hold on
parameters $\gamma$ and $h$ to ensure that $\ket{M_3}$ is an
eigenstate. It is possible to show\cite{eigenmmes} that the
parameters in \eqref{eq:H3} have to meet for all $i$'s the
following requirements:
\begin{equation}\label{eq:cond3}
\begin{array}{c}
\displaystyle \gamma_{i+1}^{xx}-\gamma_{i+1}^{yy} = h_{i}^x,
\qquad \gamma_{i+1}^{xy}+\gamma_{i+1}^{yx} = h_{i}^y, \\
\displaystyle \gamma_{i+1}^{yz}+\gamma_{i}^{zy} = 0, \qquad
\gamma_{i+1}^{xz}+\gamma_{i}^{zx} = 0, \\ \displaystyle
\sum_{i=1}^3 h_i^z = 0.
\end{array}
\end{equation}
While conditions \eqref{eq:cond3} ensure that $\ket{M_3}$ is an
eigenstate of the Hamiltonian, it is not generally possible to
analytically verify its nodegeneracy. Moreover, even when the
above conditions are satisfied, $H$ still contains non-parallel
couplings, such as $\sigma^x\sigma^y$, which can be difficult to
engineer and are not even necessary to avoid degeneracy. In order
to simplify the picture, let us choose to study a simple operator,
meeting all the requirements in Eq.~\eqref{eq:cond3} and depending
on two parameters:
\begin{equation}\label{eq:H3Jk}
H_3(J,k) = J \sum_{i=1}^3 \sigma_i^z \sigma_{i+1}^z + k
\sum_{i=1}^3 \left( \sigma_i^x \sigma_{i+1}^x - \sigma_i^x
\right).
\end{equation}
This Hamiltonian involves Ising-like couplings along the $x$ and
$z$ axes with independent and general coupling constants, and an
interaction with an external field along $x$, which is finely
tuned with the Ising coupling in the same direction. It is evident
that conditions \eqref{eq:cond3} are satisfied. The operator
$H_3(J,k)$ can be analytically diagonalized. The $J$-term gives
$\ket{M_3}$ a nonzero eigenvalue
\begin{equation}
H_3(J,k) \ket{M_3} = 3 J \ket{M_3},
\end{equation}
and avoids degeneracy with states having non aligned spins, such
as $\ket{001}$ and $\ket{110}$, while the $k$-term removes the
degeneracy with the GHZ state $(\ket{000}-\ket{111})/\sqrt{2}$,
preserving the eigenstate property of $\ket{M_3}$. The spectrum as
a function of $J/k$ is plotted in Figure \ref{fig:spectrum3}.

If $k\neq 0$, the MMES eigenstate has only two degeneracy points
in $J=0$ and $J=-k/2$, while it is always at least twofold
degenerate when $k=0$. An analysis of the spectrum in Figure
\ref{fig:spectrum3} confirms that $\ket{M_3}$ can never be the
nondegenerate ground state of the system. However, it is possible
to identify ranges in which it is the {\it nondegenerate
first-excited state}. These are
\begin{equation}
J/k < 1/2 \quad \text{for} \quad k>0
\end{equation}
and
\begin{equation}
J > 0 \quad \text{for} \quad k<0.
\end{equation}
In this case, a tentative protocol to generate the MMES state
would be to let the system relax towards the ground state, and
then drive it to the lowest excited level.

\subsection{$n=4$}

It is known that in the case of four qubits, perfect MMES
satisfying $\pi_{\mathrm{ME}}=1/2^2=1/4$ do not
exist.\cite{higuchi1,higuchi2,MMES} There is a proof\cite{wallach}
that the minimum of potential of multipartite entanglement is
instead $\pi_{\mathrm{ME}}=1/3$. In this section, we will analyze
the properties of one of the minimizers,
\begin{equation}\label{eq:M4}
\ket{M_4} = \frac{1}{4} \sum_{\ell=0}^{15} \zeta_{\ell}^{(4)}
\ket{\ell},
\end{equation}
where $\{ \ket{\ell} \}_{0\leq \ell \leq 15} = \{
\ket{0000},\ket{0001},\dots,\ket{1111} \}$ are the states of the
computational basis and
\begin{equation}
\zeta^{(4)}=\{1, 1, 1, 1, 1, 1, -1, -1, 1, -1, 1, -1, -1, 1, 1,
-1\}
\end{equation}
is the array of coefficients. It is one of the so-called uniform
real MMES, enumerated in Ref.~\refcite{multent}, whose
coefficients are all real and equal in modulus. While $\ket{M_4}$
is maximally mixed with respect to the bipartitions
$(\{1,2\},\{3,4\})$ and $(\{1,3\},\{2,4\})$, the purity of the
bipartition $(\{1,4\},\{2,3\})$ reads $\pi_{\{1,4\}}=1/2>1/4$.

For $n=4$, it has been found in Ref. \refcite{eigenmmes} that
there exist 24 independent and nontrivial elementary Hamiltonians
with $\ket{M_4}$ as an eigenstate, which are two-body and nearest
neighbor. Most of them involve non parallel couplings, like
$\sigma_i^x\sigma_{i+1}^z$, which we manage to avoid in building
the Hamiltonian $H_3(J,k)$ in the case $n=3$. On the contrary, in
the four-qubit case it is necessary to include this type of
interactions, since it is impossible for $\ket{M_4}$ to be a
nondegenerate eigestate if only parallel couplings and
interactions with external fields are present. One of the simplest
Hamiltonian which for which $\ket{M_4}$ is a generally
nondegenerate eigenstate reads
\begin{equation}\label{eq:H4Jk}
H_{4}(J,k) = J(\sigma_4^x\sigma_1^z+\sigma_3^x\sigma_2^z)+k
\left(\sigma_1^x\sigma_4^z+\sigma_2^x\sigma_3^z +
\sigma_2^x\sigma_1^z+\sigma_1^x\sigma_2^z- \sum_{i=1}^4 \sigma_i^z
\right).
\end{equation}
Notice that also an external field along the $z$ axis, finely
tuned to the two-body interaction terms, is present. As for $H_3$
in \eqref{eq:H3Jk}, the parameter $J$ determines the eigenvalue of
the MMES, through
\begin{equation}
H_4(J,k) \ket{M_4} = 2 J \ket{M_4},
\end{equation}
while terms multiplying the parameter $k$ remove a fourfold
degeneracy, related to the fact that there exist three states,
which are orthogonal to $\ket{M_4}$ and to each other, and have
the same expectation value $2J$ of the energy (see discussion in
Section 3). Incidentally, the Hamiltonian we are considering has
also a second MMES eigenstate. This is due to the fact that a
reflection of the $z$ axis maps $H_4(J,k)$ onto $-H_4(J,k)$, which
has the same eigenstates with opposite eigenvalues. Thus, the
state
\begin{equation}\label{eq:Mt4}
\ket{\tilde{M}_4} = \frac{1}{4} \sum_{\ell=0}^{15}
\zeta_{15-\ell}^{(4)} \ket{\ell}
\end{equation}
with $\zeta^{(4)}$ defined after Eq.~\eqref{eq:M4}, which is
obtained by reflecting $\ket{M_4}$ with respect to the $z$ axis,
is also an eigenstate of $H_4$, with eigenvalue $-2J$.

\begin{figure}
\centering
\includegraphics[width=0.8\textwidth]{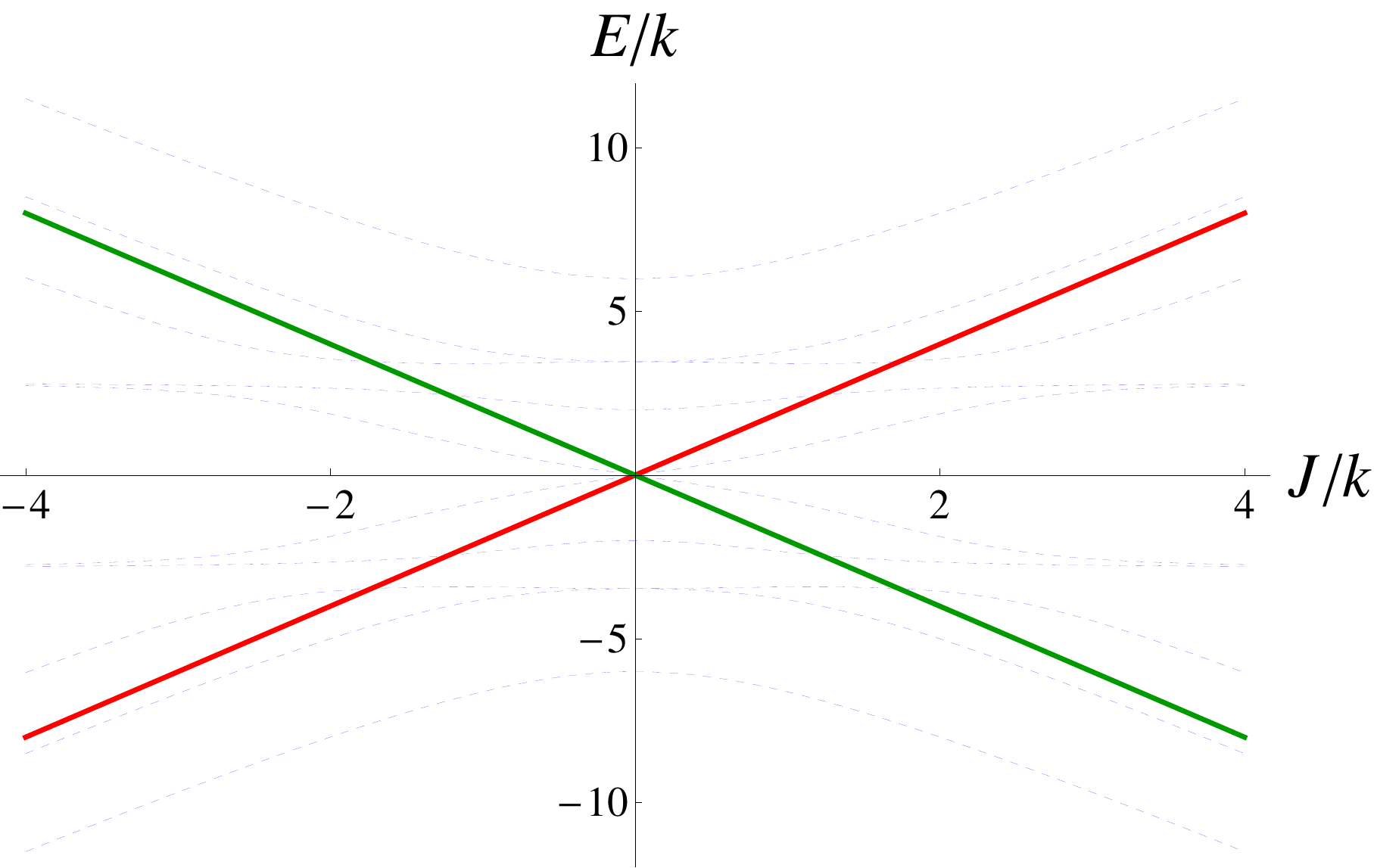}
\caption{Spectrum of the Hamiltonian $H_4(J,k)$ as a function of
$J/k$. The solid (red) positive-slope line represents the
eigenvalue $E=2J$ of $\ket{M_4}$, while the solid (green)
negative-slope line corresponds to the eigenvalue $E=-2J$ of the
MMES state $\ket{\tilde{M}_4}$ [see Eq.~\eqref{eq:Mt4} and
discussion.] The dashed (blue) lines represent the other energy
levels, with the zero eigenvalue being fourfold
degenerate.}\label{fig:spectrum4}
\end{figure}

The spectrum of $H_4$ as a function of the reduced parameter $J/k$
is represented in Figure \ref{fig:spectrum4}.
Apart from the case $k=0$, in which the MMES eigenstate is
systematically degenerate, it can be observed from Figure
\ref{fig:spectrum4} that $\ket{M_4}$ has five degeneracy points on
the $J/k$ line, placed at
\begin{equation}
\frac{J}{k} = 0, \quad \frac{J}{k} = \pm \sqrt{\frac{3}{2}}, \quad
\frac{J}{k} = \pm \sqrt{3}.
\end{equation}
It also emerges that the lowest possible level for the chosen MMES
eigenstate is the second-excited one, which is occupied by
$\ket{M_4}$ if
\begin{equation}
k>0 \text{ and } J/k<-\sqrt{3/2} \qquad \text{or} \qquad k<0
\text{ and } J/k > \sqrt{3/2}.
\end{equation}
The impossibility to reach even the first excited state in the
case of $\ket{M_4}$ is closely related to the energetic
competition with (linearly independent) states having the same
expectation value of $H_4$, which are in this case three, while in
the $n=3$ case it was only one.

\subsection{$n=5$}

Perfect MMES do exist for a system of five qubits. Let us consider
as in the case $n=4$ an example of a uniform and real MMES, namely
\begin{equation}\label{eq:M5}
\ket{M_5} = \frac{1}{2\sqrt{2}} \sum_{\ell=0}^{31}
\zeta_{\ell}^{(5)} \ket{\ell},
\end{equation}
with
\begin{eqnarray}
\zeta^{(5)} &=&\{1, 1, 1, 1, 1, -1, -1, 1, 1, -1, -1, 1, 1, 1, 1,1, 1,1,\nonumber\\
& &-1, -1, 1, \ -1, 1, -1, -1, 1, -1, 1, -1, -1, 1, 1\}.
\end{eqnarray}
The difficulty in finding a Hamiltonian with a MMES eigenstate
evidently increase as one tackles the case $n=5$. Indeed, even if
the dimension of the Hilbert space is twice that of the previous
case, the number of nontrivial elementary terms with $\ket{M_5}$
as an eigenstate remains 24, as in the case $n=4$. It is even a
harder task to avoid degeneracy and to find Hamiltonians with the
energy of $\ket{M_5}$ close to the minimum\cite{eigenmmes}. The
reason of such a difficulty lies in the fact that, in this case,
there exist 31 states, all orthogonal to $\ket{M_5}$ and to each
other, sharing the same expectation value for {\it all} two-body
operators.

An example of Hamiltonian with a generally nondegenerate MMES
eigenstate, which can be at least numerically diagonalized, reads
\begin{eqnarray}
H_5 (J,k) & = & J \left( \sigma_1^x \sigma_2^x + \sigma_2^x
\sigma_3^x + \sigma_4^x \sigma_5^x + \sigma_5^x \sigma_1^x +
\sigma_3^y \sigma_4^y - \sigma_3^z \sigma_4^z - 2 \sigma_1^z
\right) \nonumber \\ & + & k \left( \sigma_1^y \sigma_2^y +
\sigma_2^y \sigma_3^y + \sigma_4^y \sigma_5^y + \sigma_2^z
\sigma_3^z + \sigma_4^z \sigma_5^z + \sigma_2^x  - \sigma_3^x +
\sigma_4^x - \sigma_5^x - \sigma_5^z \right). \nonumber \\
\end{eqnarray}
The eigenstate $\ket{M_5}$ has vanishing eigenvalue. Degeneracy
occurs for
\begin{equation}
k=0, \quad J=0, \quad \frac{J}{k} \simeq \pm 0.24, \pm 0.54, \pm
1.35.
\end{equation}
Since $H_5$ has roughly the same number of positive and negative
eigenvalues, the eigenstate $\ket{M_5}$ is forced to stay at the
center of the energy band. The best one can do in lowering its
energy is to choose $J/k$ in the open intervals $(0.24,0.54)$ if
$k>0$ or $(-0.54,-0.24)$ if $k<0$. In these cases, $\ket{M_5}$ is
the thirteenth excited state.

\section{(Rather non)local Hamiltonians with GHZ eigenstates}

Let us recall the definition \eqref{eq:ghz} of
Greenberger-Horne-Zeilinger states for a system of $n$ qubits:
\begin{equation}
|G_{\pm}^{n}\rangle=\frac{1}{\sqrt{2}} \left(|0\rangle^{\otimes
n}\pm |1\rangle^{\otimes n}\right). \nonumber
\end{equation}
The great homogeneity of GHZ states as $n$ varies, both in their
definition and in their entanglement properties, suggests that the
results obtained for the $n=3$ MMES $\ket{M_3}=\ket{G_+^3}$ admit
some generalization to an arbitrary number of qubits. Indeed,
general results were obtained in Ref.~\refcite{ghz}, where
necessary and sufficient conditions are discussed for a GHZ
nondegenerate eigenstate of a Hamiltonian with $k$-body couplings,
with $k<n$.

Here, we shall simply mention the main result: {\it if a
Hamiltonian does not involve terms coupling at least $[(n+1)/2]$
qubits, then the GHZ state $\ket{G_+^n}$ cannot be a nondegenerate
eigenstate.} The same result holds true for $\ket{G_-^n}$ and all
the states obtained by applying local unitary transformations to
$\ket{G_+^n}$. This limitation is frustrating for the task we are
trying to accomplish, since it implies, for example, that for
$n=5$ and $n=6$ three-body interactions are {\it necessary} to
avoid the degeneracy of a GHZ eigenstate. Anyway, one can try to
build Hamiltonians which are as local as possible, compatibly with
the above negative result, namely which contain no more than
$[(n+1)/2]$-body interaction terms.

The states $\ket{G_+^n}$ and $\ket{G_-^n}$ in \eqref{eq:ghz} are
twofold degenerate ground states of an Ising ferromagnetic
Hamiltonian, acting on the $n$ qubits on a circle:
\begin{equation}\label{eq:ising}
H_0=-\sum_{j=1}^n \sigma_j^z\sigma_{j+1}^z \quad (\textrm{with} \;
\bm{\sigma}_{n+1}\equiv\bm{\sigma}_1).
\end{equation}
In order to lift the degeneracy of the ground state, a suitable
perturbation should be added to $H_0$, in a way to preserve the
eigenstate property of $\ket{G_+^n}$. The ``minimal'' perturbation
consists in two terms, with a proper relative sign, flipping
complementary sets of spins. In order to keep the range of
coupling as small as possible, one can choose terms which flip
$[n/2]$ and $n-[n/2]$ spins. If one uses $\sigma^x$ matrices to
flip spins, a suitable Hamiltonian reads
\begin{equation}\label{eq:Hlambda}
H(\lambda) = H_0 + \lambda H_1,
\end{equation}
with
\begin{equation}
H_1 = \sigma_1^x \sigma_2^x \dots \sigma_{[n/2]}^x -
\sigma_{[n/2]+1}^x \dots \sigma_n^x.
\end{equation}
The relative minus sign ensures that $\ket{G_+^n}$ is an
eigenstate for all $\lambda$, being
\begin{equation}
H_1\ket{G_+^n}=0.
\end{equation}
On the other hand, $H_1$ couples $\ket{G_-^n}$ to another linear
combination of two states with opposite spins, namely
\begin{equation}
\ket{\tilde{G}_-^n} = \frac{1}{\sqrt{2}} \left( \ket{0_1\dots
0_{[n/2]}1_{[n/2]+1}\dots 1_n} - \ket{1_1\dots
1_{[n/2]}0_{[n/2]+1}\dots 0_n} \right).
\end{equation}
The states $\ket{G_-^n}$ and $\ket{\tilde{G}_-^n}$ span a
two-dimensional invariant sector of $H_1$. More generally, all the
states which are superpositions with zero relative phase of two
vectors of the computational basis with opposite spins are
eigenstates of $H(\lambda)$, while superpositions with relative
phase $\pi$ are coupled in two-dimensional sectors.

\begin{figure}
\centering
\includegraphics[width=0.8\textwidth]{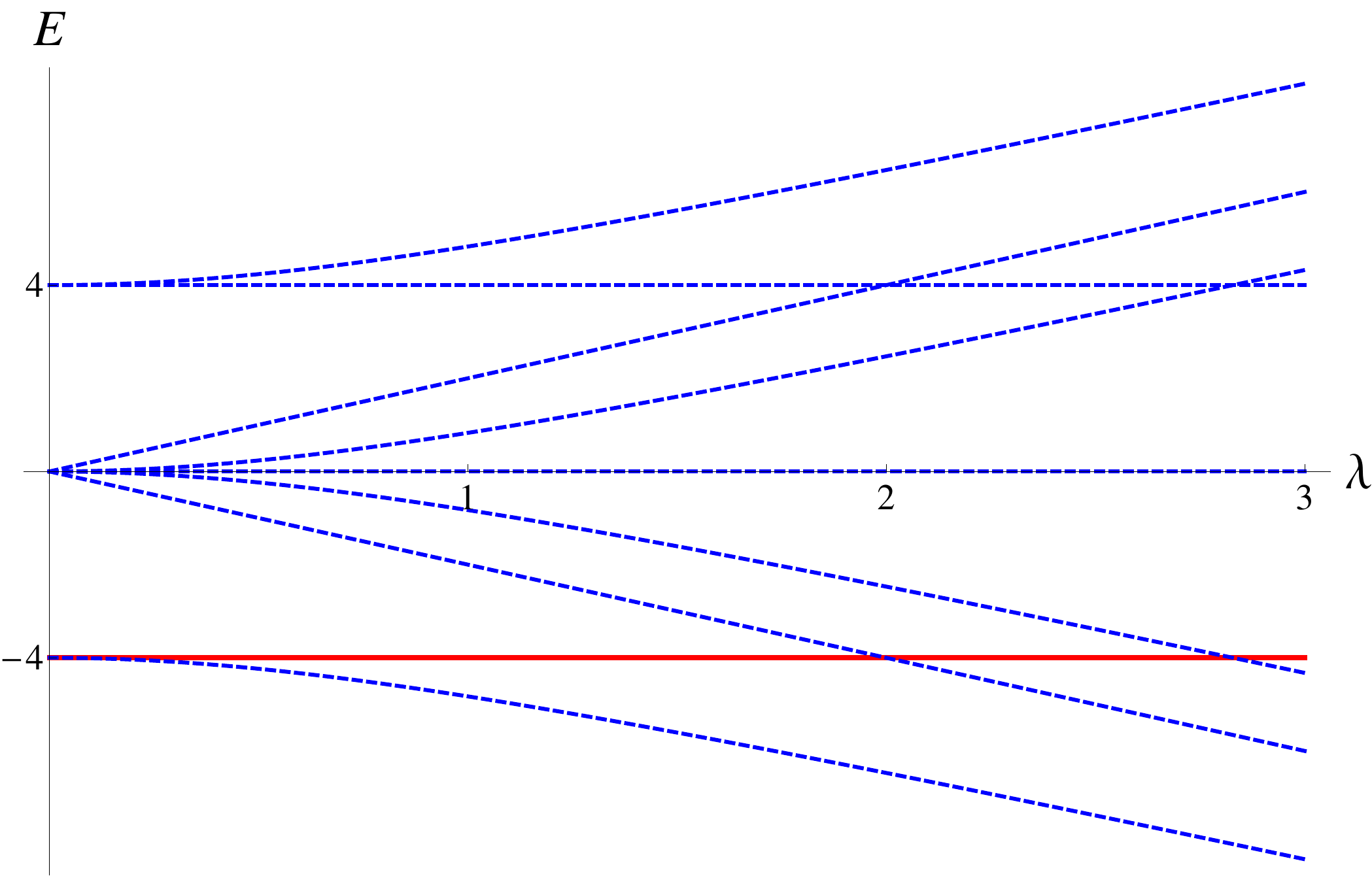}
\caption{Energy levels of $H(\lambda)$ for $n=4$. The solid (red)
line represents the constant energy of the eigenstate
$\ket{G_+^n}$. The dashed (blue) lines correspond to other energy
levels, some of which are degenerate. Notice the crossings between
the GHZ energy and excited levels at $\lambda=2$ and $\lambda=2
\sqrt{2}$, and the degeneracy with the ground state energy level
at $\lambda=0$. Eigenvalues are even functions of
$\lambda$.}\label{fig:spectrumGHZ}
\end{figure}

Diagonalization for arbitrary $n$ is performed in
Ref.~\refcite{ghz}. In Figure \ref{fig:spectrumGHZ}, the full
spectrum is plotted for the relevant case $n=4$, which is the
largest system in which a two-body perturbation is sufficient to
remove degeneracy.
The ground state of the system is a linear combination of
$\ket{G_-^n}$ and $\ket{\tilde{G}_-^n}$, with energy
\begin{equation}
E_0 = -n - 2 \left( \sqrt{ 1+ \lambda^2 } -1 \right).
\end{equation}
This implies that $\ket{G_+^n}$, whose energy is $-n$, can be at
most the first-excited state, as expected. Indeed, it is the
nondegenerate first-excited state in the ranges $-2<\lambda<0$ and
$0<\lambda<2$. Accidental degeneracies for the eigenstate
$\ket{G_+^n}$ occur for
\begin{equation}
\lambda = \pm 2 k \qquad \text{ and } \qquad \lambda = \pm 2
\sqrt{k(k+1)}
\end{equation}
with $k=0,1,\dots,[n/2]$, including the case in which it is a
degenerate ground state for $\lambda=0$. These results show that,
even if it is necessary to include (generally) nonlocal terms in
the Hamiltonian to lift the degeneracy between $\ket{G_+^n}$ and
$\ket{G_-^n}$, it is indeed sufficient to add a {\it small}
nonlocal perturbation $\lambda H_1$, with $\lambda\ll 1$, to
obtain $\ket{G_+^n}$ as a nondegenerate first-excited state.

\section{Conclusions and outlook}

We have reviewed results on the possibility to obtain $n$-qubit
MMES (for $n\leq 5$) and GHZ states as eigenstates of Hamiltonians
involving only local interaction terms, and analyzed the
properties of case-study Hamiltonians with multipartite-entangled
eigenstates. Since both MMES and GHZ states exhibit very
distributed non-local correlations, the answer to this problem is
nontrivial. Indeed, we found that these states can never be
nondegenerate ground states of local Hamiltonians for $n>2$. We
also showed that this limitation does not arise in the case of W
states. In the Hamiltonians we study throughout the paper, we find
that, already for $n=3$, MMES can at most be the first
non-degenerate excited state, and the situation worsens as $n$
increases. We find severe limitations on the possibility to find a
few-body Hamiltonian with a GHZ nondegenerate eigenstate. All
these results can be interpreted as a manifestation of
entanglement frustration.\cite{nielsen1,nielsen2,NJP}

Besides the foundational interest, research on the possibility to
encode a multipartite-entangled state into the eigenstate of a
local Hamiltonian is also motivated by few-qubit applications.
Practically realizable quantum tasks usually involve only a very
small number of qubits (see e.g.~Ref.~\refcite{zoller}.) The most
advanced quantum applications require that the qubits be prepared
in highly entangled states with large fidelity. One expects that
more performing applications would become possible by making use
of MMES: some examples were proposed in Refs.~\refcite{Adesso} and
\refcite{FFLMP}. This clearly requires efficient methods to
prepare MMES with large yield and efficiency. For example, if a
MMES were the ground state of some Hamiltonian $H$, then one could
engineer it by constructing $H$ and letting the system relax
toward its ground state. However, since this is not the case, one
must consider a partial relaxation combined with control
techniques or devise alternative strategies. Another possible
mechanism for generating MMES is dynamical rather than static.
This is obviously related to the degree of complexity of a quantum
circuit that generates MMES using, for instance, two-qubit gates.
An interesting matter of investigation would be to analyze and
compare the efficiency and reliability of the two strategies.

It would also be interesting to study the entangling power of the
Hamiltonians with MMES and GHZ
eigenstates,\cite{zanardi,scott,caves}. The performance in
entangling an initial state (e.g.~a factorized one) can be
compared with the ones of standard Hamiltonians such as Ising- or
Heisenberg-type, or local Hamiltonians with randomly distributed
parameters. Future work will be devoted to this topic.

\section*{Acknowledgments}

The author thanks Paolo Facchi, Giuseppe Florio and Saverio
Pascazio for fruitful interaction in the research on this topic
and for useful suggestions on the present paper.

\end{document}